# V1405 Cas Nova 2021: About the nature of a multi-maxima nova.

Erik Wischnewski[1]

[1]Bundesdeutsche Arbeitsgemeinschaft für Veränderliche Sterne e.V., Berlin, Germany

24 November 2021

**ABSTRACT**

*Context:* The Nova V1405 Cas, which erupted on 18 March 2021, showed a pre-maximum phase of almost two months. After the main maximum, over a period of several months, the nova only weakened slightly and showed clear fluctuations in brightness.

*Aims:* The aim of this work is to investigate the relationship between brightness and half-width (FWHM) and equivalent width (EW) of the Hα emission line, and from this to infer possible causes for the abnormal behavior of the Nova.

*Methods:* Magnitudes and spectra from online databases were used, the Hα line of the spectra was measured and 2-day normal values were formed in order to establish a temporal comparison.

*Results:* After the main maximum, during the period investigated by JD 2 459 292 to 2 459 305 at intervals of 2–5 weeks, secondary maxima with different characteristics resulted. The maxima of the equivalent width were 6–14 days after the respective brightness maxima. The maxima are interpreted as subsequent outbreaks of the nova. Possibly these are subsequent weaker eruptions of the nova or other mechanisms that lead to an increase in density in the envelope.

**Key words**  Nova – V1405 Cas – light curves – equivalent width

# Introduction

## Characteristics of the Nova

On 18 March 2021 at 10:10 UT, Yuri Nakamura discovered the Nova Cassiopeiae 2021. The star was already known as a close binary and eclipsing star of the type *W Ursae Majoris* (EW) under the following names:

CzeV3217
UCAC4 756-077930
PNV J23244760+66111140
Gaia EDR3 2015451512907540480

The equatorial coordinates (J2000.0) are:

Right ascension = $23^h\,24^m\,48^s$
Declination = $+61°\,11'\,15"$

The distance is d = $1693^{+74}_{-67}$ pc [6].

A period of 0.1883907 days and a brightness of V = 15.6 mag [1] respectively 14.87–14.96 mag [7] are given for the pro-outburst light changes of the star. The orbital period of the double star is 0.376938 days [7].

After the eruption as the nova on 18 March 2021 (RJD 292.3), the star was given the designation *V1405 Cas* and was classified as a slow nova (NB).[1]

## Aim of the investigation

After the nova initially appeared as a classic He/N nova [11], it did not show the expected decrease in brightness. Instead, the brightness increased again very strongly after a few weeks. The nova remained bright for months and also highly variable. It also turned into a FeII nova.

The last nova investigated in more detail by the author was V339 Del (2013), which was a classic FeII nova and for which physical models could also be developed. It is completely different from this nova. On the basis of the brightness measurements as well as the full width at half maximum (FWHM) and equivalent width (EW) of the Hα emission line, an attempt should be made to find explanatory approaches.

# Data

Because of the high brightness, the striking behavior of the light curve and the proximity to the open star cluster Messier 52, which is only 0.4° north of the nova, the star was very often observed photometrically and spectroscopically by amateurs. On 18 October 2021, 48 630 brightness information could be downloaded from the *AAVSO International Database* (AID).

| band | total | used |
|---|---|---|
| visual | 3 147 | |
| CV | 32 405 | 32 298 |
| CR | 3 | |
| U | 5 | |
| B | 1 209 | 1 107 |
| V | 5 161 | 4 877 |
| R | 1 165 | 1 130 |
| I | 1 139 | 1 075 |
| TB | 701 | 614 |
| TG | 3 307 | 633 |
| TR | 388 | 385 |

**Table 1** Number of downloaded and used data from the total of 48 630 data sets. *Source: AAVSO International Database, 18 October 2021.*

---

[1] For the sake of simplicity and readability, the Julian date JD is written as the revised JD (RJD), where: RJD = JD − 2 459 000.

Only the UBVRI magnitudes should be included in the evaluation. However, because the number was too low, U could not be taken into account. In addition to BVR, the similar colors TB, TG and TR of the RGB channels of a digital camera with Bayer matrix (tricolor) were used and compared with the BVR magnitudes according to Johnson-Kron-Cousins. If in this work the magnitudes R and I are used, then $R_C$ and $I_C$ of the Kron-Cousins-System are meant. The CV magnitudes were only loaded for the purpose of comparison with the V magnitudes, but not used for the evaluation of the nova.

After checking the reliability of the data, only the number of measurements in the third column of Table 1 were included in the evaluation. The significantly lower number of measurements used in TG can be explained, among other things, by the fact that observers sometimes indicated only 1–2 mmag as an error. As this is very unlikely even under the most favorable data reduction conditions, these values were not trusted. The individual examination of one observer showed that his values were initially in the main field, but later turned out to be a half magnitude brighter (→ Figure 3). This reinforces the suspicion about the usability, so that only about 20% of the TG were used.

Only V and CV magnitudes were considered, the errors of which are below 0.05 mag. For all other color bands, 0.1 mag was used as the margin of error.

71 spectra were used to measure the H$\alpha$ emission line:

| Number of spectra | |
|---|---|
| Source | Number |
| AVSPEC (AAVSO) | 57 |
| BAA | 8 |
| ARAS | 6 |

**Table 2** Number of AAVSO, BAA and ARAS spectra used.

# Methods

On the one hand, in order to be able to compare the different color bands with one another and to form color indices, normal values were formed. For this purpose, all data were averaged over a progressive 2-day interval. The first interval comprises the measurements from RJD 292.0–294.0 and is assigned to the date RJD 293.0. The next interval then comprises RJD 294.0–296.0 etc. At the same time, this smooths the light curve.

In this work a total of 71 spectra with a spectral resolution of R ≥ 500 were evaluated. Of these, 26 have a resolution of R ≥ 15 000. The equivalent width (EW) and the full width at half maximum (FWHM) of the H$\alpha$ emission line were measured. Further spectra between R ≥ 200 and R ≤ 500 were unusable for various reasons.

# Results

## Preliminary examination of the brightness values

**Differences |** Figure 4 to Figure 6 show the differences between the standard magnitudes B, V and R and the tricolor magnitudes TB, TG and TR and Figure 7 contains the difference V–CV. Further information on the comparisons is given in the appendix.

**Color indices |** Figure 8 to Figure 13 show the color indices of the standard magnitudes. Notes are also given in the appendix. The diagrams show no peculiarities or systematic trends, with the exception of the color indices (B–R) and (V–R), which have a significant increase at the time of the first maximum of the H$\alpha$ equivalent width. This can still be recognized rudimentarily in the color indices with (V–I), (B–I) and (R–I). This corresponds to the expectations, since the H$\alpha$ line lies in the red band.

For the sake of completeness, the three color indices were also visualized with the addition of the tricolor magnitudes (→ Figure 14 to Figure 16). In Figure 16, in addition to the first maximum, the second maximum at RJD 437 is clearly evident.

**State diagrams |** The three color magnitude diagrams in Figure 17 to Figure 19 do not show any abnormalities and, from the current point of view, do not provide any information on the physics of the nova. The same applies to the two-color index diagram in Figure 20.

## Light curve

Shortly after the discovery, the nova reached its maximum with a visual magnitude of 7.5 mag (≈RJD 294). In the period RJD 300–315 it reached an elongated intermediate minimum that fluctuated around 8.0 mag. During this period the spectrum of the nova changed from a He/N nova to a FeII nova.

The visual magnitude maximum was V = 5.2 mag on 10/11 May 2021 (RJD 345.5 ± 0.5), and thus 53.2 days after the outbreak of the nova.

| Maxima von V | |
|---|---|
| Maximum (RJD) | mag |
| **2021-05-11 (346)** | **5.3** |
| 2021-06-07 (373) | 6.9 |
| 2021-06-17 (383) | 6.9 |
| 2021-06-29 (395) | 6.8 |
| **2021-07-27 (423)** | **6.0** |
| 2021-09-09 (467) | 6.6 |
| 2021-09-21 (479) | 6.6 |

**Table 3** Maxima of the V magnitudes, given in brackets is the RJD based on JD 2 459 000. The column ›mag‹ shows the approximate magnitudes. (Note: Because of the smoothing, the maximum date deviates by about 0.1 mag from the calculated regression value.)

In contrast to a classic nova, in which the main maximum occurs approx. 2–3 days after the pre-maximum, it took almost two months for the Nova Cas 2021 (≈ 52 days).

## Hα emission line

**Half width** | The full width at half maximum (FWHM) of the Hα emission line in Figure 23 was measured consistently in the interval 6510–6616 Å. The measurements of the high-resolution spectra lie in the general stray area of all measurements. In addition to local fluctuations, which in particular have two maxima around RJD 357 and RJD 437, the entire course has an increasing trend, that is, the rate of expansion increases on average over months. This fact is remarkable, which is why it will be discussed in more detail later.

**Equivalent width** | Two maxima at the equivalent width of the emission line of Hα appear particularly clearly in Figure 24. An overview of all minima and maxima is given in Table 4.

| Minima and Maxima of $EW_{H\alpha}$ | | |
|---|---|---|
| Minimum | Maximum | EW [Å] |
| 2021-05-10 (345) | | 30 |
| | **2021-05-22 (357)** | **1 700** |
| 2021-06-05 (371) | | 350 |
| | 2021-06-13 (379) | 940 |
| 2021-06-18 (384) | | 480 |
| | 2021-06-24 (390) | 870 |
| 2021-07-02 (398) | | 350 |
| | 2021-07-09 (405) | 870 |
| 2021-07-20 (416) | | 210 |
| | **2021-08-10 (437)** | **2 150** |
| 2021-09-18 (476) | | 70 |

**Table 4**  Minima and maxima of the equivalent width of the Hα emission line, the RJD is given in brackets based on JD 2 459 000.

**Instrinsic line flow** | Figure 25 shows the intrinsic Hα line flow, calculated on the basis of the visual magnitude V as

$$\frac{-EW_{H\alpha}}{10^{0.4 \cdot V}}.$$

Note: The minus sign is only intended to compensate for the negative sign of the equivalent width for emission lines.

In this case, intrinsic means that the line flow has been freed (cleared) of fluctuations in the continuum radiation, the continuum being only approximately represented by the photometric magnitude V.

The two maxima are also reflected in the intrinsic line flow, although weaker than in the case of the equivalent width. In addition, the R magnitude was used to represent the continuum, in which the maxima are even more clearly developed. This is to be expected since the R magnitude includes the Hα lines.

## Parameters of the star

**Extinction** | The interstellar absorption $A_V$ is calculated from

$$A_V = 3.2 \cdot E_{B-V}.$$

With $E_{B-V} = 0.55$ mag [12] the interstellar absorption is $A_V = 1.76$ mag.

A rough calculation with an average extinction in the surrounding area of the sun of 1.0 mag/kpc results in a good agreement of $A_V = 1.7$ mag.

**Absolute magnitude** | The distance modulus is calculated from

$$m - M = 5 \cdot \lg d - 5 + A_V$$

at the time of the maximum (m = 5.2 mag) and the distance (d = 1690 pc) to m–M = 12.90 mag, corresponding to an absolute magnitude of $M_V = -7.7$ mag.

**Luminosity** | The estimation of the luminosity L is based on the assumption that the bolometric correction $BC_V$ of the nova is approximately equal to that of the sun, via the absolute visual magnitude $M_V$. From

$$\left(\frac{L}{L_\odot}\right) = 10^{0.4 \cdot (M_{V,\odot} - M_V)}$$

with $M_{V,\odot} = 4.8$ mag, the maximum luminosity of the nova is $L \approx 90\,000\ L_\odot$ (→ Figure 26).

**Radius** | The radius can be derived from the Stefan-Boltzmann law:

$$\frac{L}{L_\odot} = \left(\frac{R}{R_\odot}\right)^2 \cdot \left(\frac{T}{T_\odot}\right)^4$$

With a photospherical envelope temperature of the nova of T = 9000 K [16] and the effective temperature of the sun of $T_\odot = 5778$ K, a maximum radius of $R \approx 125\ R_\odot$ is calculated (→ Figure 27).

**Distance of the double star** | From the Kepler relationship

$$\left(\frac{a}{AU}\right)^3 = \left(\frac{M_1 + M_2}{M_\odot}\right) \cdot \left(\frac{U}{year}\right)^2$$

With U = 0.377 d ≈ 0.001 years [7] and the assumed masses $M_1 = 1\ M_\odot$ and $M_2 = 2\ M_\odot$ a major semi-axis of a = 0.0147 AE ≈ 3.2 $R_\odot$ results. The nova envelope therefore covered the entire binary system.

# Discussion

On 3 April 2021 (RJD 308) the nova showed a helium spectrum; on 19 April 2021 (RJD 324) the iron lines were already clearly in the foreground [13]. As a result, the initial He/N nova must have mutated into an FeII nova in the period from 5 April 2021 to 20 April 2021 (RJD 310–325). This is exactly the broad minimum between the pre-maximum and maximum of the brightness (→ Figure 21). There seems to be a connection between the course of brightness (light curve) and the change in the nova spectral type. On 8 May 2021 the spectrum at the maximum brightness corresponded to a textbook FeII nova [14].

Highly ionized emission lines in the spectra indicate that the nova photosphere has not expanded enough to cool itself down [11].

The spectrum looks very similar to that of the nova V339 Del. The complete absence of forbidden metallic emission lines suggests that even 15 weeks after the eruption and 55 days after maximum brightness ($\approx$ RJD 400) the density of the ejection is high enough to prevent the formation of these lines [15].

After a first outbreak with an increase in brightness of $\approx$ 8.1 mag (= pre-maximum), a second one followed with a further increase in brightness of $\approx$ 2.3 mag (= main maximum) and then further weaker increases.

A closer look at the temporal development of the brightness on the one hand and the equivalent width on the other hand shows that the EW maxima occurred 6-14 days after the respective brightness maxima.

Several mechanisms are conceivable as the cause, but in the opinion of the author they must all result in an increase in the matter density of the shell. It is not yet possible to deduce from these data whether these occurred through eruptive eruptions or in a more gentle way.

As shown in the section ›Instrinsic line flow‹, an increase in the equivalent width can also be caused by an increase in the continuum radiation. The timely increase in the V magnitude and the equivalent width EW indicate precisely this fact. The instrinsic line flow is only partially corrected for continuum, as Figure 25 shows, where the maxima are still present.

The high matter densities of the photosphere (shell) mentioned in [11] and [15] can be explained by the additional, more or less strong material replenishment, which can be determined both in the light curve and in the equivalent width.

**Expansion velocity** | In a classic nova, the matter of the eruption is accelerated by the radiation pressure. Nevertheless, one observes a decreasing rate of expansion over time, because it relates to the photosphere. The photosphere, however, migrates inward over time, because the shell becomes thinner and thinner due to the lack of material supply. This allows us to see deeper and deeper layers of the envelope, which has not yet been accelerated that much [10].

In the case of Nova Cas 2021, however, a steadily increasing (average) rate of expansion can be observed in the first 7 months. this can only be explained if matter is constantly being supplied over the entire period. Apparently this does not happen continuously, but in bursts, as the maxima in the brightness and the flow of lines show.

# Conclusions

V1405 Cas initially presented itself as a He/N nova in the pre-maximum and then mutated into a classic FeII nova during the pre-maximum phase.

This pre-maximum phase, which usually lasted only a few days, lasted for more than seven weeks. Detailed spectral studies will be necessary to understand the exact process during this time.

After the main maximum, the nova not only remained relatively bright for months, but also regularly showed outbreak-like secondary maxima at intervals of 2–5 weeks. These maxima first appeared in the visual continuum and 1–2 weeks later in the H$_\alpha$ equivalent width. The author suspects that a similar process is responsible for this as well as for the long-lasting pre-maximum phase. There is still a need to explain the time lag between the continuum and line maximum.

Since the nova outbreak is caused by the accreted matter from the companion (donor), it cannot be completely ruled out that the cause of the special behavior of this nova lies in this accretion flow. In the case of a recurring nova of the type NR it takes 10–10 000 years until the next eruption; here it is only a few weeks. So there has to be a crucial difference between the two ›re-excitation mechanisms‹, or it is simply a completely different process.

In this investigation, the light curve and the flow of lines were compared with each other for the first time and the nova was described as a quasi multi-maxima nova. Building on this, further data must now be collected as already described.

# Acknowledgements

This research made use of the SIMBAD database and of the VizieR catalogue access tool, operated at CDS, Strasbourg, France, and the International Variable Star Index (VSX) database, operated at AAVSO, Cambridge, Massachusetts, USA.

The author acknowledges with thanks the variable star observations from the AAVSO International Database contributed by observers worldwide and used in this research [2].

The author thankfully acknowledges further the spectra from the AAVSO Database AVSPEC [3], ARAS Spectal Database [4] and BAA Spectroscopy Database [5] contributed by Hugh Allen, Mariusz Bayer, Christophe Boussin, David Boyd, John Briol, Erik Bryssinck, Mario Clemens, Scott Donnell, James Foster, Joan Guarro Fló, Peter Somogyi and Tim Stone, and used in this research.

The auther would like to thank Tom Fields for providing and supporting the RSpec software [8].

# Appendix

The appendix contains all diagrams and figures mentioned in the text.

## Environment map

The close proximity to Messier 52 and the relatively bright stars in the neighborhood allow the nova to be found quickly.

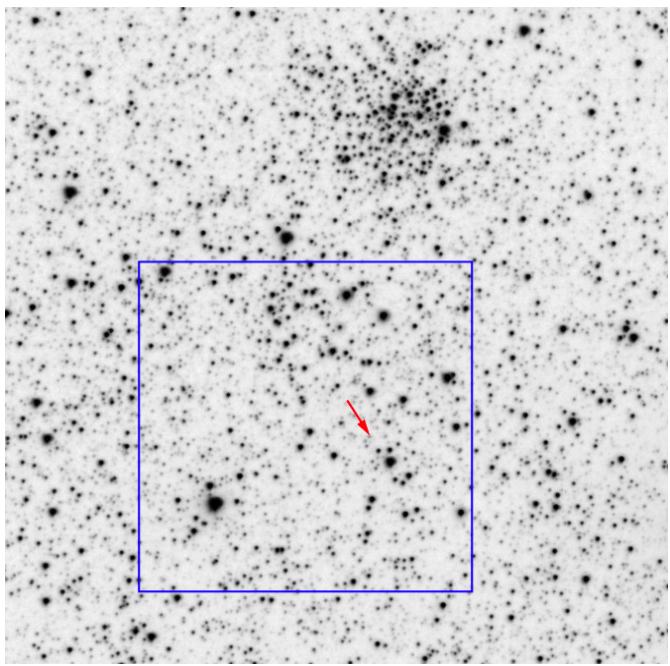

**Figure 1**  Area map with M 52 from 2013-09-29. The progenitor of the nova is marked (red arrow). The blue frame indicates the approximate section of Figure 2.

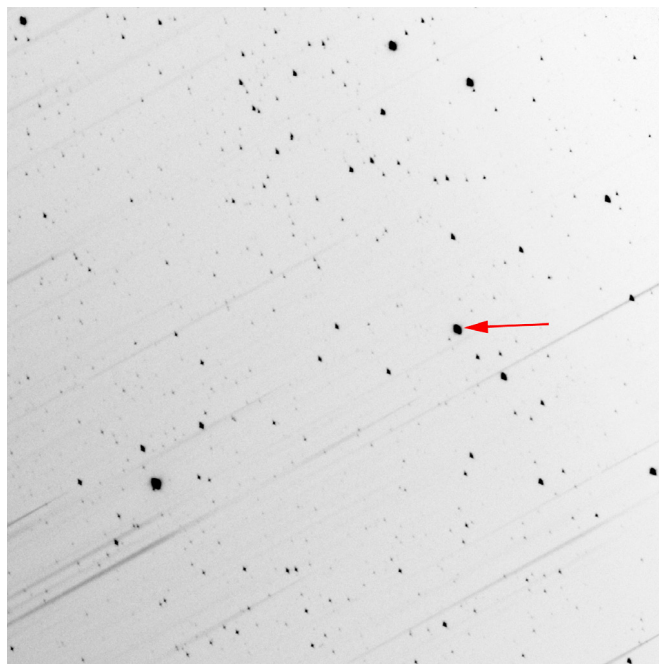

**Figure 2**  Part of the spectral recording with clearly recognizable nova (red arrow).

## Plausibility check

The observations were subjected to a plausibility check before they were used for evaluation.

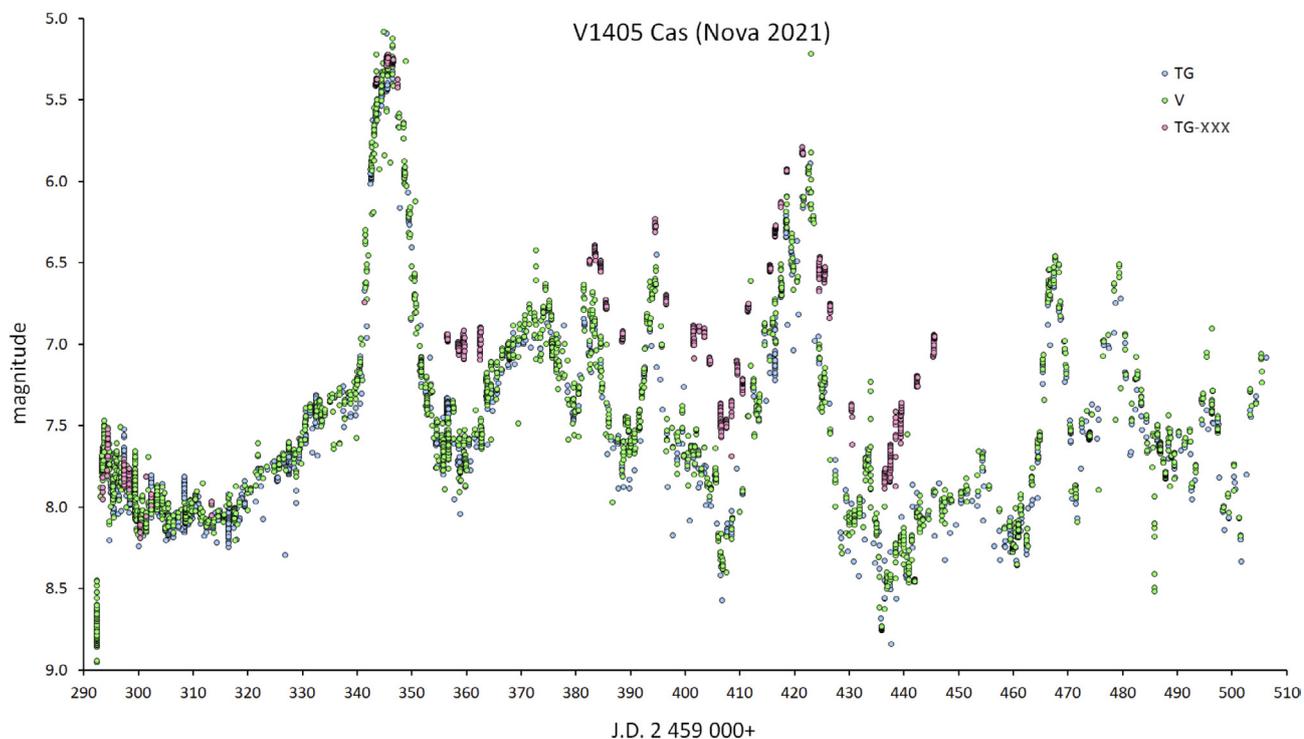

**Figure 3**  Example of the measurements in the TG band of the (anonymized) observer XXX. The clear deviations in the period JD 2 459 350 to JD 2 459 450 led, among other things, to not taking this data into account (for the time being). The observer has been informed.

## Comparison of tricolor to standard magnitudes

Since the data of the bands TB, TG and TR of a tricolor digital camera with Bayer matrix are typically not calibrated by the observers to the photometric standard system according to Johnson-Kron-Cousins, it was necessary to check how these values relate to the standard magnitudes B, V and R behavior. The differences are considered.

For further evaluations, the standard and tricolor magnitudes are used together and designated as:

   B'= [B, TB]
   V'= [V, TG]
   R'= [R, TR]

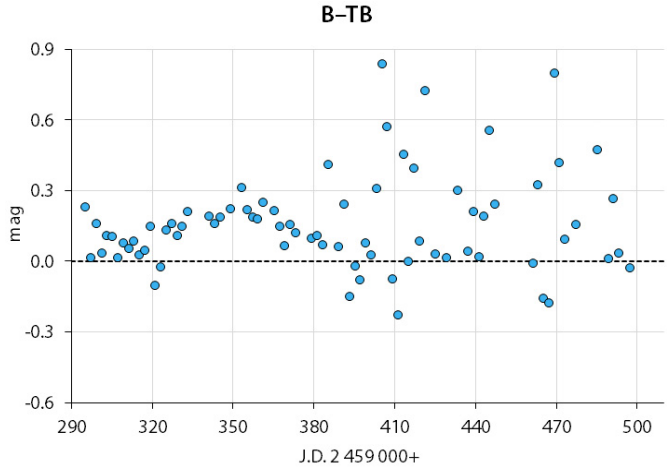

**Figure 4**   Difference between the Johnson magnitude B and the tricolor magnitude TB. It is noticeable that the TB magnitudes are predominantly brighter than the B magnitudes and also scatter strongly.

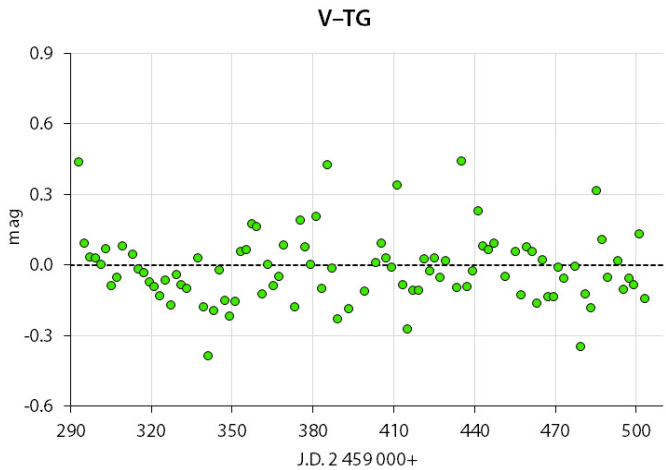

**Figure 5**   Difference between the Johnson magnitude V and the tricolor magnitude TG (G = green). The TG magnitudes are at the same level as the V magnitudes, but with a clear spread.

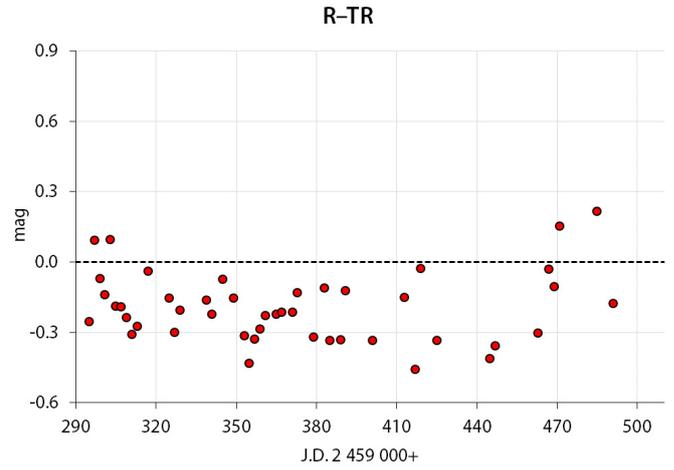

**Figure 6**   Difference between the Johnson-Kron-Cousins magnitude R ($R_C$) and the tricolor magnitude TR. It is noticeable that the TR magnitudes are predominantly darker than the $R_C$ magnitudes and also scatter significantly.

## Compare V to CV

Monochromatic CCD astronomical cameras, which only use a UV/IR blocking filter, record the entire spectral range from blue to red (referred to as *clear*). If the V magnitudes are used for the reference stars, this brightness is referred to as CV. A comparison with magnitudes, which were also recorded with a Johnson V filter, shows a clear deviation with a large scatter. For this reason the CV values were not used.

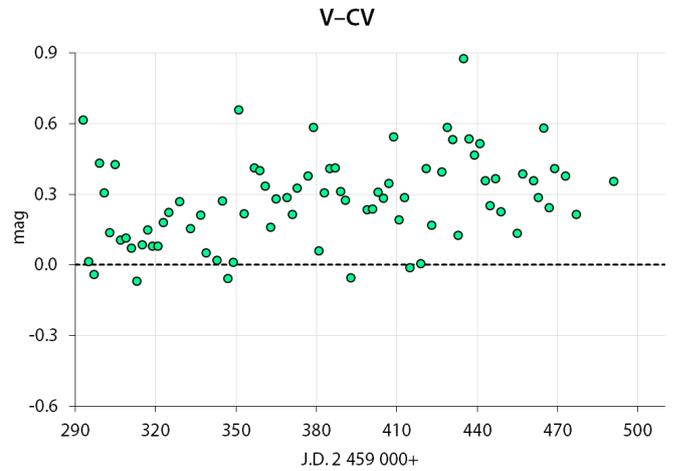

**Figure 7**   Difference between the Johnson magnitude V and the unfiltered magnitude CV. The CV magnitudes are generally brighter than the V magnitudes and also scatter very strongly.

## Color indices of the standard magnitudes

In addition to the light curve in a certain color band, the light curves of the color indices, i.e. the difference between two color bands, are also of interest. A physical interpretation is not yet possible in the case of the nova, only the color indices B–R and V–R reflect the first maximum of the equivalent width of the H$\alpha$ line.

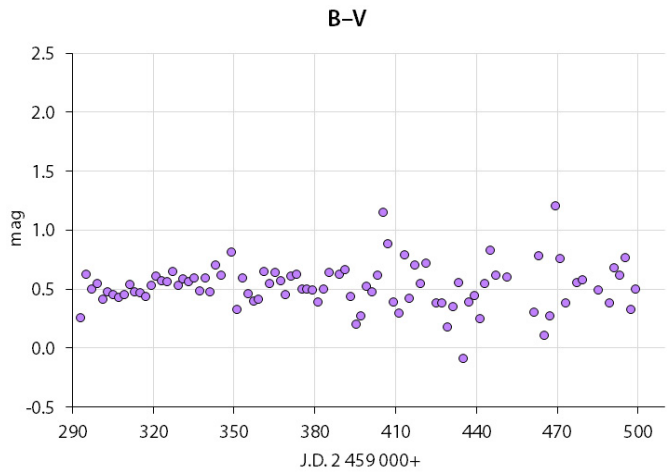

**Figure 8**   Color index B–V of the normal values over 2 days.

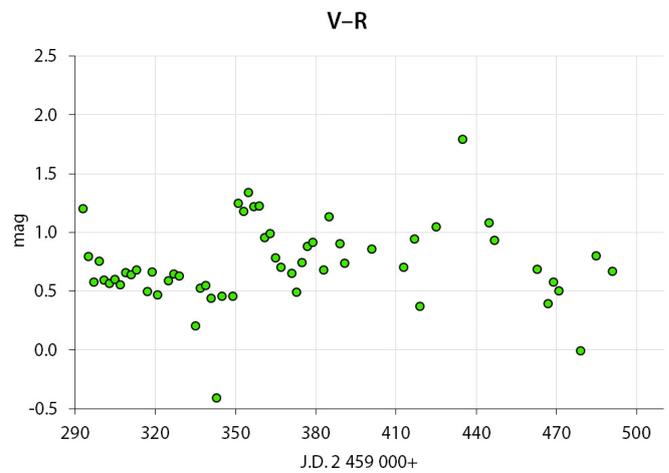

**Figure 11**   Color index V–R of the normal values over 2 days.

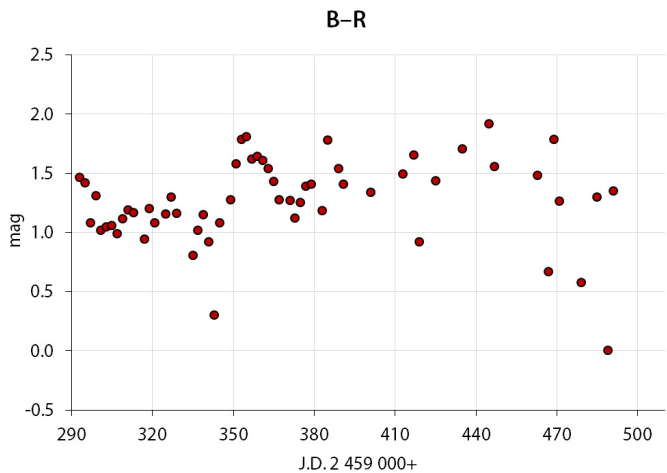

**Figure 9**   Color index B–R of the normal values over 2 days.

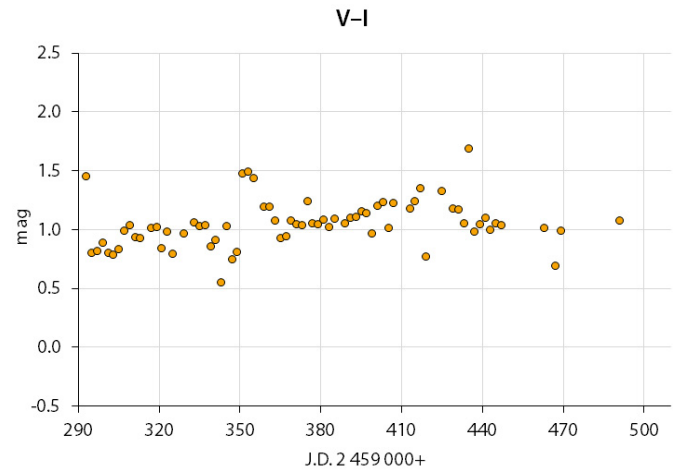

**Figure 12**   Color index V–I of the normal values over 2 days.

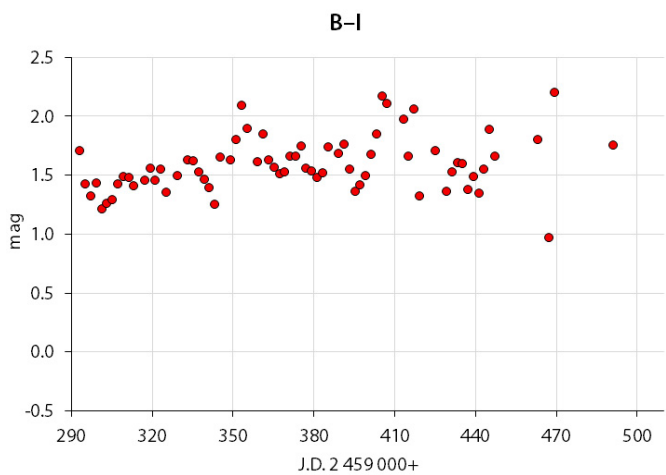

**Figure 10**   Color index B–I of the normal values over 2 days.

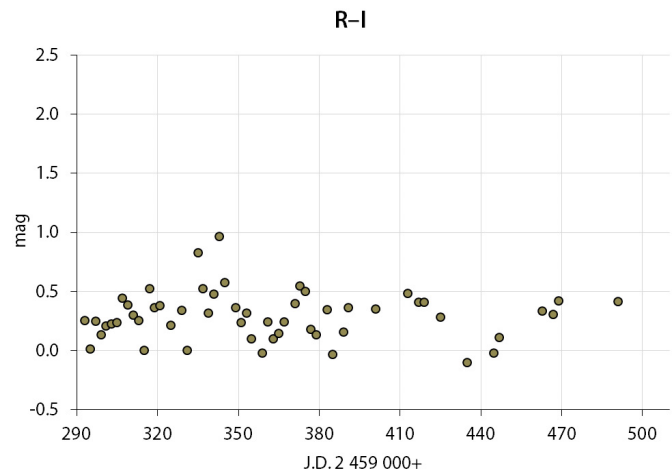

**Figure 13**   Color index R–I of the normal values over 2 days.

## Color indices of the combined magnitudes

The inclusion of the tricolor measurements in the generation of the normal values reduced the scatter of the color indices slightly.

## State diagrams

The three following color magnitude diagrams and the two-color index diagram do not show any particular conspicuousness that would be worth mentioning.

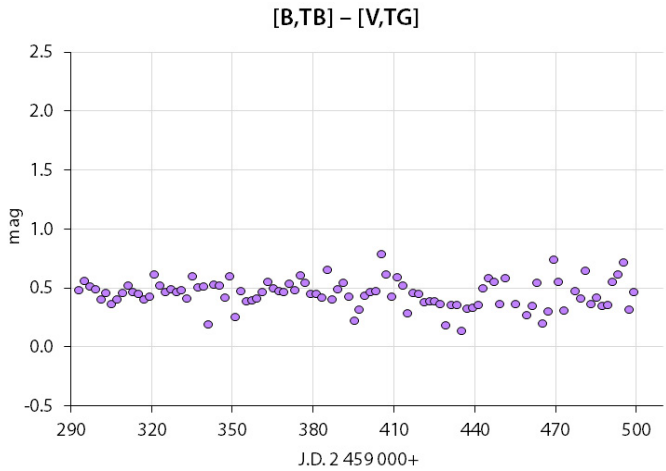

**Figure 14**  Color index (B'–V ') of the normal values over 2 days including the TB and TG values.

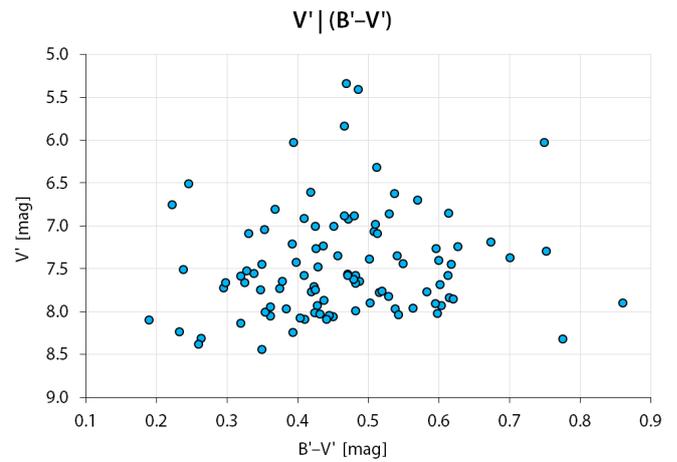

**Figure 17**  Visual magnitude V' versus the color index (B'– V').

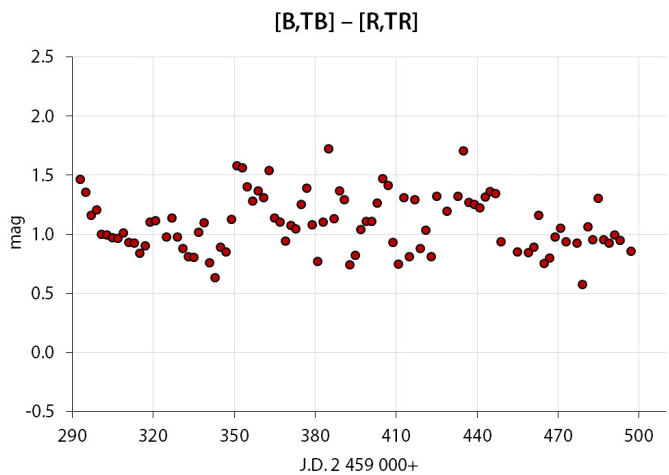

**Figure 15**  Color index (B'–R ') of the normal values over 2 days including the TB and TR values.

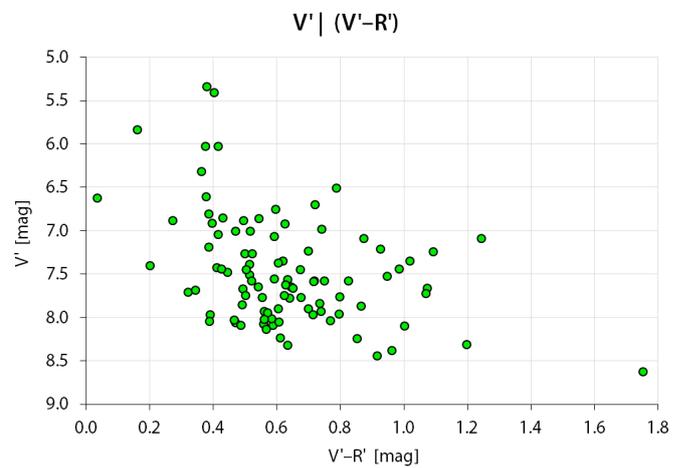

**Figure 18**  Visual magnitude V' versus the color index (V'-R').

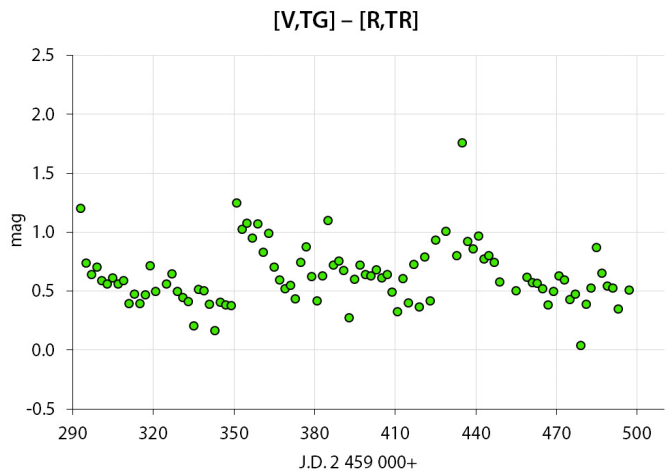

**Figure 16**  Color index (V'–R ') of the normal values over 2 days including the TG and TR values.

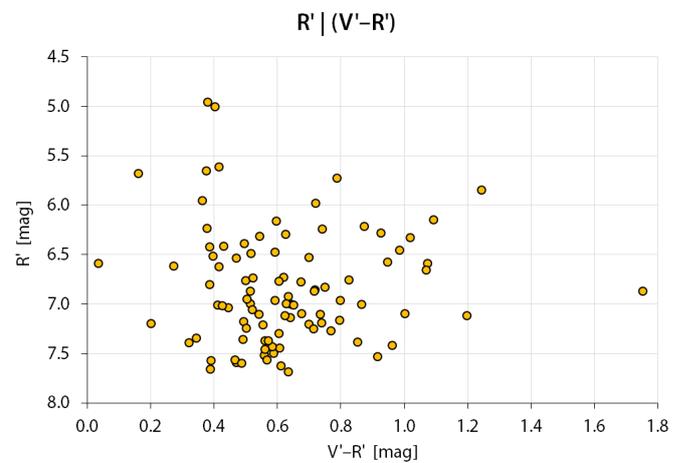

**Figure 19**  Red magnitude R' versus the color index (V'– R').

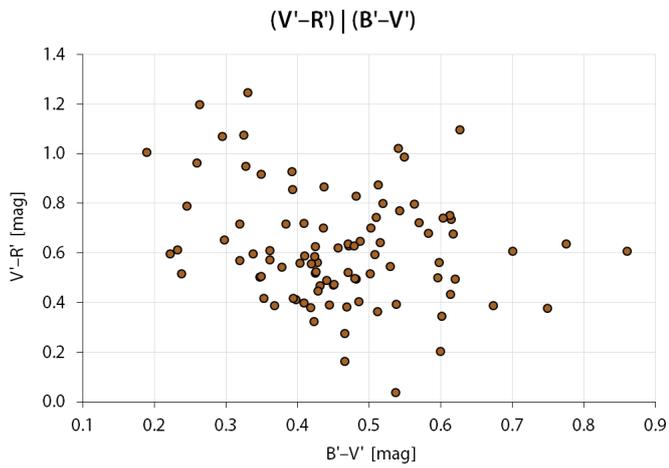

**Figure 20**  Two-color index diagram of the color indices (V'–R ') and (B'–V').

## Light curve, Hα emission line, luminosity and radius

The discussion of the following diagrams takes place in the main part of this paper.

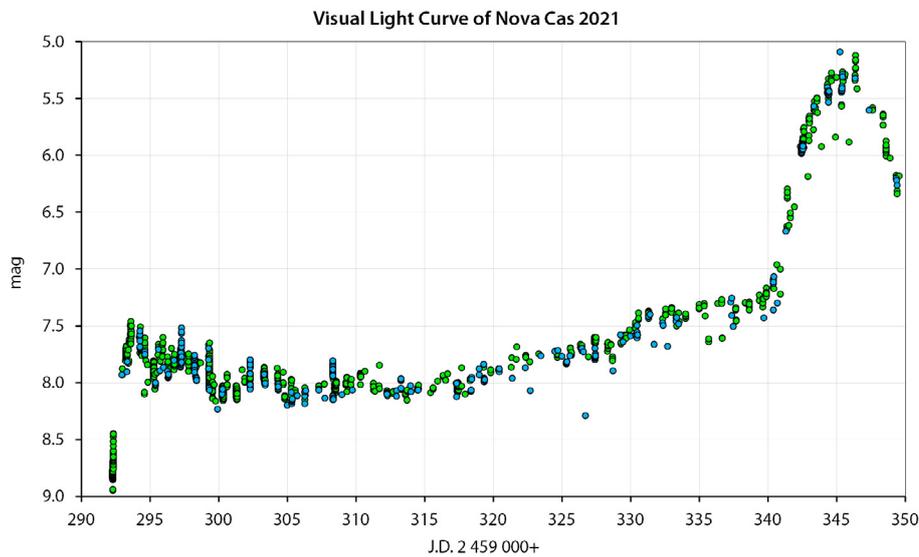

**Figure 21**
Individual observations of the first two months in the bands V (green dots) and TG (blue dots).

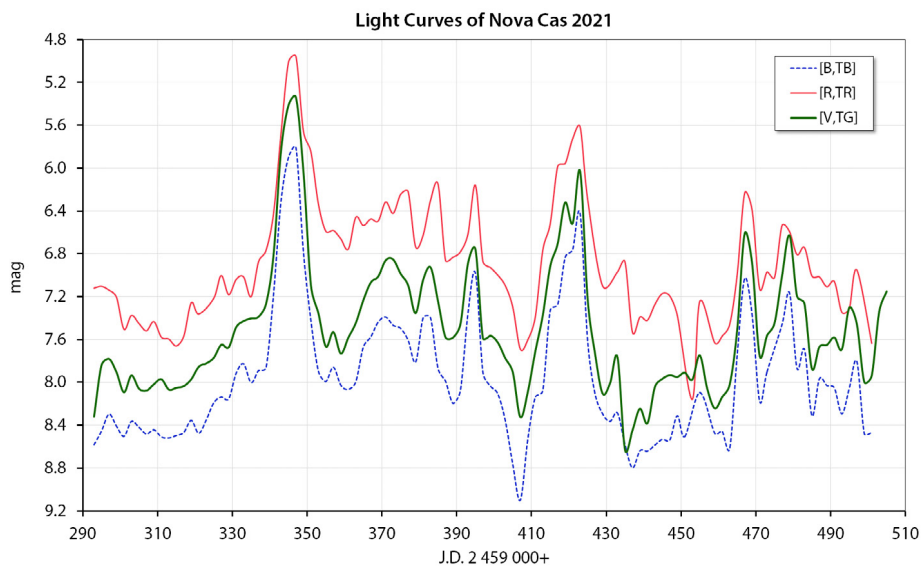

**Figure 22**
Light curves of the 2-day normal values in the three color ranges blue, green and red, each using the standard magnitudes B, V and R as well as the tricolor magnitudes TB, TG and TR (= B', V', R').

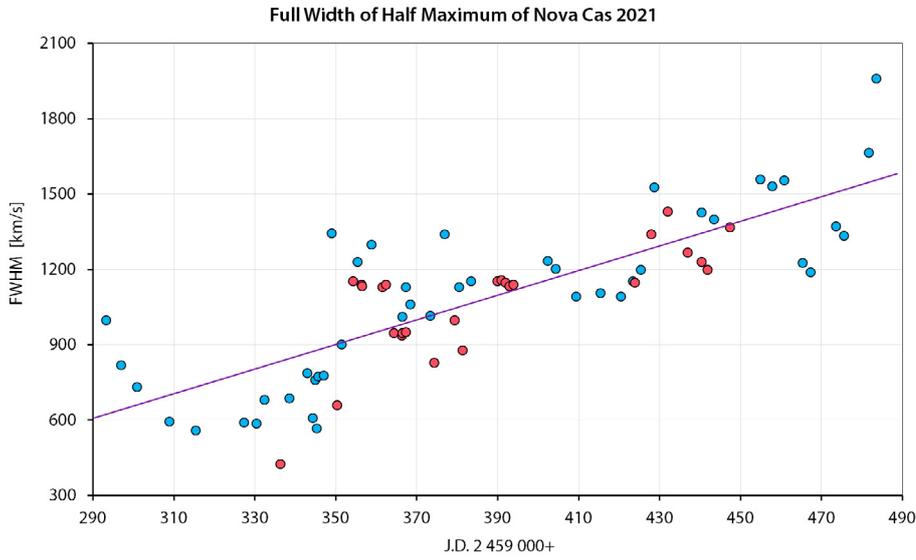

**Figure 23**
Full width at half maximum (FWHM) of the H$_\alpha$ line, given as speed in km/s. The blue dots indicate spectra with R = 500–1500, the red dots represent spectra with R ≥ 15000. The expansion rate increases steadily on average, superimposed by temporary fluctuations.

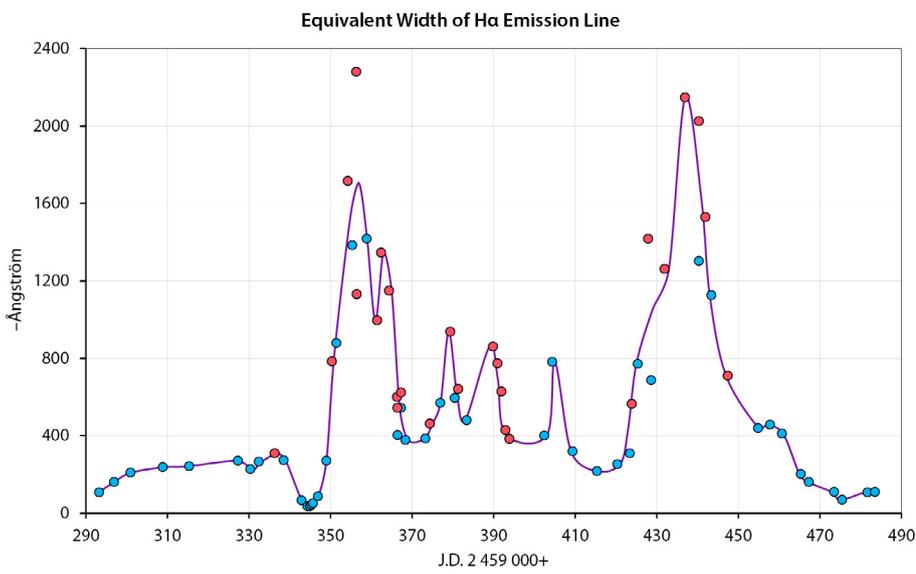

**Figure 24**
Equivalent width (EW) of the H$_\alpha$ line, one (averaged) value per 2-day interval, given in Ångström. The blue dots indicate spectra with R = 500–1500, the red dots represent spectra with R ≥ 15000. Two dots were averaged for the purple curve.

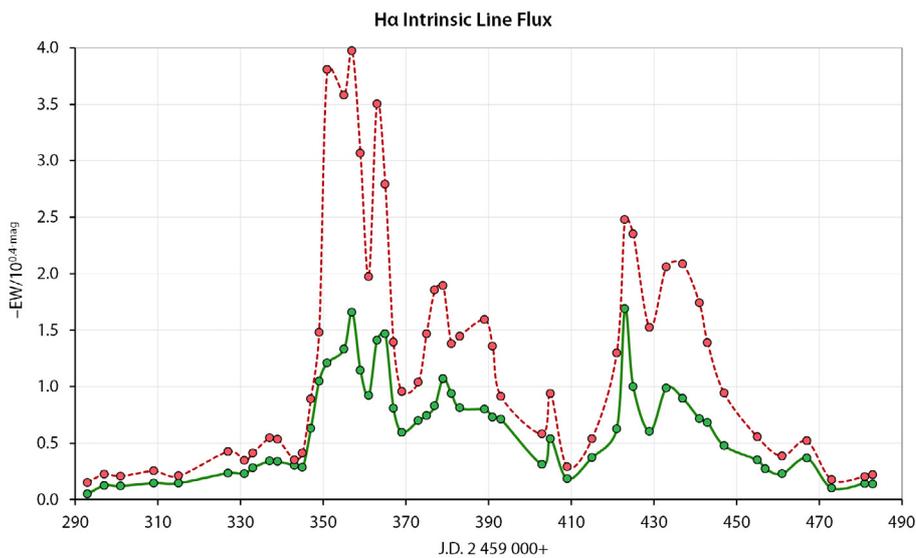

**Figure 25**
Intrinsic line flow of the H$_\alpha$ line, one (averaged) value per 2-day interval, using the magnitudes V' (green dots) and R' (red dots).

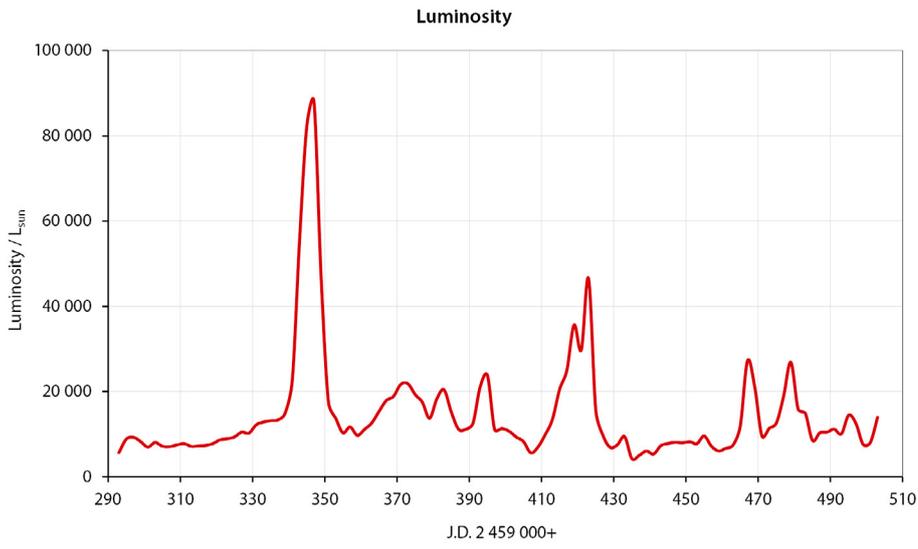

**Figure 26**
Luminosity in units of solar luminosity.

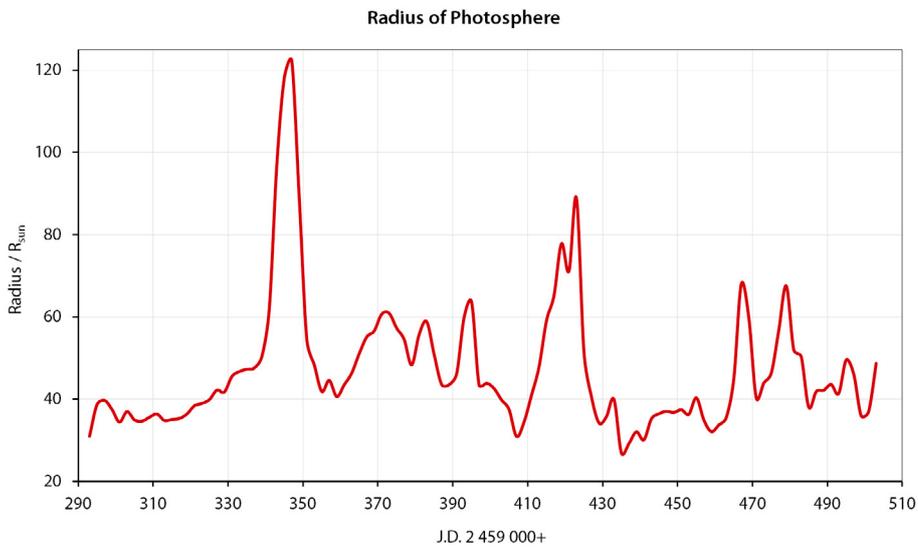

**Figure 27**
Radius of the photosphere of the shell in units of the solar radius.

12